# Experimental Demonstration of Sixth Order Degenerate Band Edge in Coupled Microstrip Waveguides

Farshad Yazdi, Dmitry Oshmarin, Tarek Mealy, Ahmad T. Almutawa, Alireza Nikzamir, and Filippo Capolino

Department of Electrical Engineering and Computer Science, University of California, Irvine, CA 92697, USA

***Abstract*—We show the physical realization and experimental demonstration of an exceptional point of 6th order degeneracy in a triple ladder (or three-way) microwave waveguide realized using three coupled microstrips on a grounded dielectric substrate. This three-way waveguide supports six Bloch eigenmodes and all coalesce onto a degenerate single eigenmode at a given frequency. The three-way waveguide is gainless, and this exceptional point is associated to a vanishing group velocity and its multiple derivatives. Indeed the $\omega - k$ dispersion diagram, that we call 6th order degenerate band edge (6DBE) has six coalescing branches. We provide the experimental verification of a 6th order exceptional point by evaluating the degenerate wavenumber-frequency dispersion diagram from the measurement of scattering parameters of a six-port unit cell. We also show the resonant behavior of a cavity made of the three-way waveguide with finite length. The unique properties of 6DBE can be exploited in designing innovative high-*Q* resonators, oscillators, filters, and pulse shaping devices.**

## I. Introduction

Degeneracy of order six in electromagnetic waveguides means that six eigenmodes coalesce and form a single degenerate mode. Here, we imply that the degeneracy is in both *wavenumbers* and *polarization states* of the six Bloch eigenmodes in a waveguide, forming a 6th order degenerate band edge (6DBE).

A particular class of exceptional points of degeneracy (EPD) in periodic structures is known as a degenerate band edge (DBE) where *four* eigenmodes in a periodic passive, lossless, waveguide coincide at the band-edge [1]–[5]. Note that the presence of the DBE and the 6DBE, discussed here, do not require the presence of losses and gain in the system. The fourth-order DBE has been shown in periodic layered media [1], periodic transmission lines [6], [7], metallic [3] and optical waveguides [8]–[10]. Experimental demonstration of the DBE has been shown in microstrip technology [5], [11], in a circular metallic waveguide [12], and in an optical waveguide [9]. A strong resonance has been shown experimentally in Ref. [13] using a variation of the DBE, called split band edge. In Ref. [6], a double ladder periodic microstrip waveguide was introduced that exhibits a DBE, and in Ref. [5], it was shown that such a structure exhibits higher quality factor and stability advantages associated with DBE resonance. In Ref. [14], implementation of a three-way partially coupled microstrip waveguide using lumped elements was presented, demonstrating a stationary inflection point (SIP) associated to a real wavenumber, which is a special EPD of order three realizable in lossless waveguides. The experimental demonstration of the SIP was shown in [15].

Here, we demonstrate theoretically and experimentally the existence of a sixth-order degeneracy (6DBE) in the Bloch wavenumber-frequency dispersion relation for a periodic waveguide implemented by three coupled microstrip lines. This condition happens when six Floquet-Bloch modes (eigenstates) coincide at the center of the Brillouin zone, here intended with the wavenumber interval of $(0, 2\pi/d)$ where $d$ is the period of the waveguide. At a 6DBE, the modal Floquet-Bloch dispersion is characterized by $(\omega_d - \omega) \propto (k - k_d)^6$ where $k$ is the Floquet-Bloch wavenumber, $\omega$ is the angular frequency, $k_d = \pi/d$ is the 6DBE wavenumber (at the center of the Brillouin zone), and $\omega_d$ is the angular frequency at which the 6DBE occurs. The exponent 6 indicates the sixth order degeneracy, implying that not only the group velocity $v_g = \partial\omega/\partial k$ of the Floquet-Bloch mode vanishes at the 6DBE, but also $\partial^n\omega/\partial k^n = 0$ for $n = 1$ to 5, while $\partial^6\omega/\partial k^6 \neq 0$.

We provide a simple implementation of the 6DBE in a three-way waveguide made of three coupled microstrips over a grounded dielectric susbstrate (Fig. 1). The unit cell of the three-way mictrostrip is a 6-port network that is an extension of the 4-port unit cell circuit of lumped elements in Ref. [16] and of the 4-port unit cell microstrip line in Ref. [5]. Implementation in other waveguide technologies involving a three-way structure (i.e., three coupled waveguides) is possible as well. Note that in general terms of idealized propagation based on coupled mode theory, a sixth-order degeneracy was already discussed in Ref. [8]. Furthermore, the 6DBE was already shown via transmission line (TL) simulations in a coupled-resonator optical waveguide (CROW), together with a possible application as a low-threshold laser [17]. This paper provides the experimental demonstration of the occurrence of the 6DBE in a waveguide at microwave frequencies and show the resonance behavior in a cavity made of a finite-length structure.

In Section II, the geometry of the unit cell of the periodic microstrip waveguide supporting a 6DBE is provided where the degeneracy behavior is shown in the modal dispersion diagram of the infinitely long, periodic, lossless structure using a microstrip waveguide transmission-line model implemented in well known commercial software packages. In Section III, the resonance behavior of the periodic three-way waveguide with finite length is shown. In Section IV, the experimental verification of the 6DBE existence in the microstrip waveguide is provided by carrying out scattering parameters measurements on a single periodic unit cell, where both full-wave simulations and measurements are in a very good agreement. The occurrence of the 6DBE is also demonstrated by showing the vanishing of the coalescence parameter, i.e., a parameter that measures the hyperdistance among the 6 eigenvectors in the waveguide system.

## II. Exceptional Degeneracy of 6<sup>th</sup> Order

To obtain a 6DBE in *reciprocal* waveguide structures, a coupling between at least three modes, for example in the three microstrip waveguides shown in Fig. 1, is required. This waveguide allows six modes to exist, three modes in each direction. The six modes coalesce into a single eigenmode at the 6DBE frequency $\omega_d$ , by resorting to proper coupling and symmetry breaking in the three periodic waveguides. In Fig. 1(a), the proposed unit cell of such a three-way periodic waveguide is shown with a red-dashed line. The structure has period of $d$ = 30 mm. For its characterization, we define six microstrips electromagnetic "ports.". We consider first a lossless structure made of microstrip coupled transmission lines that are implemented on a grounded dielectric substrate) with a thickness of 0.508 mm (20 mils), and a dielectric constant of 3.

To analyze and calculate the eigenmodes of the periodic waveguide, we use a transfer matrix formalism, which is discussed in detail in Refs. [18], [19]. We define a six-dimensional state-vector $\mathbf{\Psi}(z)=\left[\mathbf{V}^T(z) \quad \mathbf{I}^T(z)\right]^T$ to describe the evolution of the eigenmodes, where $T$ denotes the transpose operation, $\mathbf{V}(z)=[V_1,V_2,V_3]^T$ and $\mathbf{I}(z)=[I_1,I_2,I_3]^T$ are the vectors that represent the voltage (referred to the ground) and current in each of the three microstrips [20]. The evolution of this state vector from a coordinate $z_1$ to $z_2$ is then described by $\mathbf{\Psi}(z_2)=\underline{\mathbf{T}}(z_2,z_1)\mathbf{\Psi}(z_1)$ , in which $\underline{\mathbf{T}}$ is the $6\times 6$ transfer matrix [18]. The results shown in Fig. 1(b) are calculated based the transfer matrix $\underline{\mathbf{T}}$ that is obatined from circuit simulation where we simulate one unit cell by cascading several microstrip equation-based blocks. Full-wave simulations are instead used to generate the results shown in Fig. 2 and Fig. 3 where the transfer matrix $\underline{\mathbf{T}}$ is obatined by transforming the $6\times 6$ $S$-parameters matrix for the 6-port unit cell into a transfer matrix. The six Floquet-Bloch eigenmodes in the periodic three-way waveguide satisfy $\mathbf{\Psi}(z+d)=e^{-jkd}\mathbf{\Psi}(z)$ where $d$ is the length of the unit cell shown in Fig. 1(a). The six wavenumbers are obtained by solving the eigenvalue problem

$$\det\left[\underline{\mathbf{T}}(z+d,z)-e^{-jkd}\underline{\mathbf{1}}\right]=0\,, \tag{1}$$

where $\underline{\mathbf{1}}$ is the $6\times 6$ identity matrix, and we have implicitly assumed the time convention $\exp(j\omega t)$ . The dispersion diagram, which is the relation between frequency and the complex-valued Bloch wavenumbers, in a lossless and infinitely long periodic waveguide is depicted in Fig. 1(b) ), calculated based on microstrip waveguide transmission-line model using circuit simulator implemented in Advanced Design System (ADS) by Keysight. The dispersion diagram is obtained by evaluating the eigenvalues $e^{-jkd}$ derived from the transfer matrix of the three-way waveguide unit cell at each frequency and then converting them into Floquet-Bloch wavenumbers. At the 6DBE frequency of 2.95 GHz, one observes that six curves coalesce. We use red circles in Fig. 1(b) to show overlapping branches either in the real or imaginary part of the wavenumbers. At the 6DBE, the dispersion relation is locally characterized as $(\omega_d-\omega)\approx h(k-k_d)^6$ , where $h$ is a parameter that defines the flatness of the dispersion near the degeneracy frequency $f_d = \omega_d/(2\pi)$ = 2.95 GHz. Based on the real and imaginary branches of the dispersion diagram shown in Fig. 1(b), no mode can propagate in a frequency range just above $\omega_d$ because of a bandgap (indeed Im($k$) ≠ 0 for all six modes). In the range $0.9f_d < f < f_d$, only two modes can propagate (one in each $z$-direction), and the other four modes are evanescent since they have $\mathrm{Im}(k/k_d)\neq 0$ ; below $f = 0.9f_d$ other cutoff conditions occur but are not discussed in this paper since they do not exhibit a sixth-order degeneracy. In the three-way waveguide in Fig. 1(a), when working at a frequency near the 6DBE frequency, interesting features are observed such as "giant" resonance, enhanced quality factor and the unique energy distribution inside 6DBE cavity as we show in the next section, in analogy to what was observed in waveguides with a fourth-order DBE [1], [4], [19], [20].

The 6DBE condition is manifested when the transfer matrix is "defective," and a complete basis of eigenvectors cannot be found [21], [22]. The 6DBE is obtained by properly choosing the coupling between the three coupled waveguides (TLs), and it can be shown that the six independent eigenvectors coalesce into a single degenerate eigenvector. This degeneracy occurs if, and only if, the transfer matrix is similar to a Jordan canonical matrix form as discussed in Refs. [5], [10]. The coalescence of the six eigenvectors can be demonstrated using the coalescence parameter concept introduced in Ref. [5] for a fourth order degeneracy, and discussed next for the 6DBE, showing that the distance between the six eigenvectors decreases to a minimum value that depends on the amount of losses and tolerances in the system.

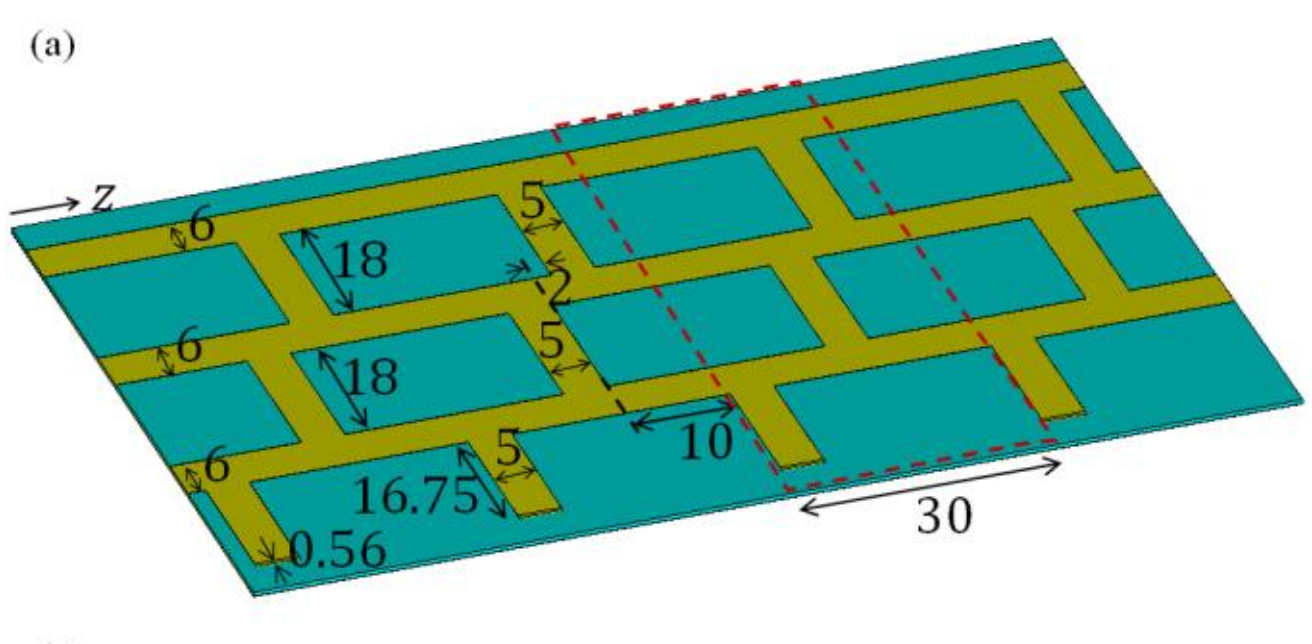

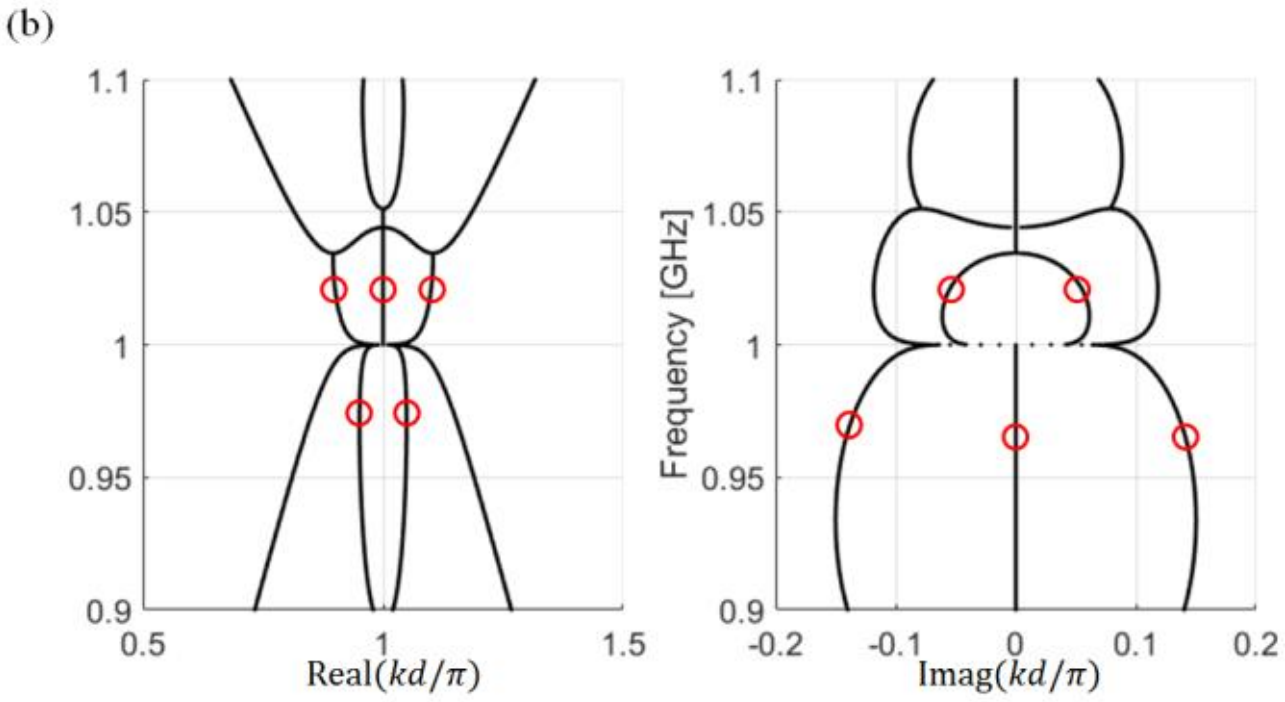

Fig. 1. (a) Periodic "three-way" waveguide, made of three coupled microstrip lines over a grounded dielectric substrate, that exhibits a sixth order DBE. The grounded substrate has a dielectric constant of 3 and the stubs are short circuited at their end. (b) Wavenumber-frequency dispersion diagram of the six Floquet-Bloch eigenmodes in the periodic three-way waveguide showing the 6DBE frequency at $f_d$ = 2.95 GHz, where six dispersion curves coalesce at the single point with wavenumber $k_d = \pi / d$ . Branches that represent two modes are denoted by a red circle. When losses are considered, the dispersion diagram is modified and shown later in Fig. 3.

## III. Resonance of Passive Cascaded Structure Including losses

We observe the resonance of a passive waveguide made of a finite number of cascaded unit cells shown in Fig. 1(a). The full-wave analysis here accounts for all losses in the dielectric, conductor, and radiation losses. We use the same substrate introduced in Section II but with a loss tangent of $\tan\delta = \varepsilon'' / \varepsilon' = 0.001$ (Rogers RO3003) with conductors (microstrip and ground plane) that aremade of copper with a thickness of 35 $\mu$m and a conductivity of $5.8\times10^7$ S/m. The 6DBE analyzed in the previous section assumed an infinitely long periodic waveguide. A resonator made of a finite-length waveguide with unit cells as in Fig. 1(a) exhibits its resonance, called 6DBE resonance, at a frequency close to $f_d$ . Such resonance exhibits some interesting and unique properties shown in Ref. [4], such as high-quality factor and its scaling with the resonator length and the distinctive energy distribution inside the finite-length cavity, which are beneficial in an oscillator, sensing and pulse shaping applications [19], [23], [24].

We study the resonance behavior by constructing a finite structure made of 8 unit-cells as shown in Fig. 2(a), with all terminals connected to a short circuit except the middle TL at the structure's left end. We assume an input phasor voltage of 1 V at the input of the resonator and observe the voltage distribution inside the resonator. Full-wave simulations are used to evaluate the $6\times6$ scattering matrix of a unit-cell, based on the the finite element frequency domain solver implemented in CST Studio Suite by DS SIMULIA. The short circuit at the end of the stub is made of a conductor via with same width of the microstrip stub and with length of 0.56 mm as shown in Fig. 1(a).

As explained earlier, the scattering matrix is then transformed into a 6×6 transfer matrix of a unit cell. The voltage distributions at the various circuit nodes are calculated by cascading the transfer matrices of the unit cells.

In Fig. 2(b) the frequency response of the voltage at the middle of the finite structure with 8 unit cells is plotted where a peak is observed for each of the three microstrip lines at a resonance frequency close to the 6DBE frequency, i.e., at $f_{r,6DBE}$ =2.988 GHz. Fig. 2(c) shows the voltage distribution inside the eight cells resonator at $f = f_{r,6DBE}$, and we see that the voltage is concentrated around the middle of the structure, in particular in the lower TL, shown in Fig. 2(a), having the highest magnitude. This unique voltage distribution implies that anything connected, like a loading resistor, at the edges of the periodic waveguide will have a minor effect on the rest of the structure. The energy distribution in a 6DBE cavity is analogous to that obtained in resonators having second or fourth order DBE [25]. This physical property has been found useful in conceiving new regimes of oscillation [26], and the voltage distribution provides the information of where an active device can be placed to have the most significant impact [23]. It also demonstrates the "slow light" effect associated with DBE, where the energy is trapped inside the cavity at a frequency near $f_d$ [25].

## IV. Experimental Verification of the 6DBE in Periodic Microstrip Circuit

The experimental verification of the existence of the sixth order EPD in the microstrip three-way waveguide shown in Fig. 1(a) is presented. We use a grounded dielectric substrate (Rogers RO3003) and conductors as in the previous section where we have discussed full-wave simulations. The fabricated unit-cell is shown in Fig. 3(a) including SMA connectors. The dispersion diagram obtained from measurement of the $6\times6$ $S$-parameters matrix, along with the full-wave simulation results (as in previous section) are shown in Fig. 3(b). They are in a good agreement, although a shift in frequency is observed due to fabrication imperfections (especially knowing exactly the lines' length) and high sensitivity of the sixth order degeneracy to perturbations. The dispersion diagram shows the coalescence of the six modes based on measurements at the degeneracy frequency $f_d = \omega_d / (2\pi)$ = 2.95 GHz. The measurement results were obtained by measuring the scattering parameters of the unit cell shown in Fig. 3(a) using a Rohde & Schwarz ZVA67 Vector Network Analyzer (VNA).

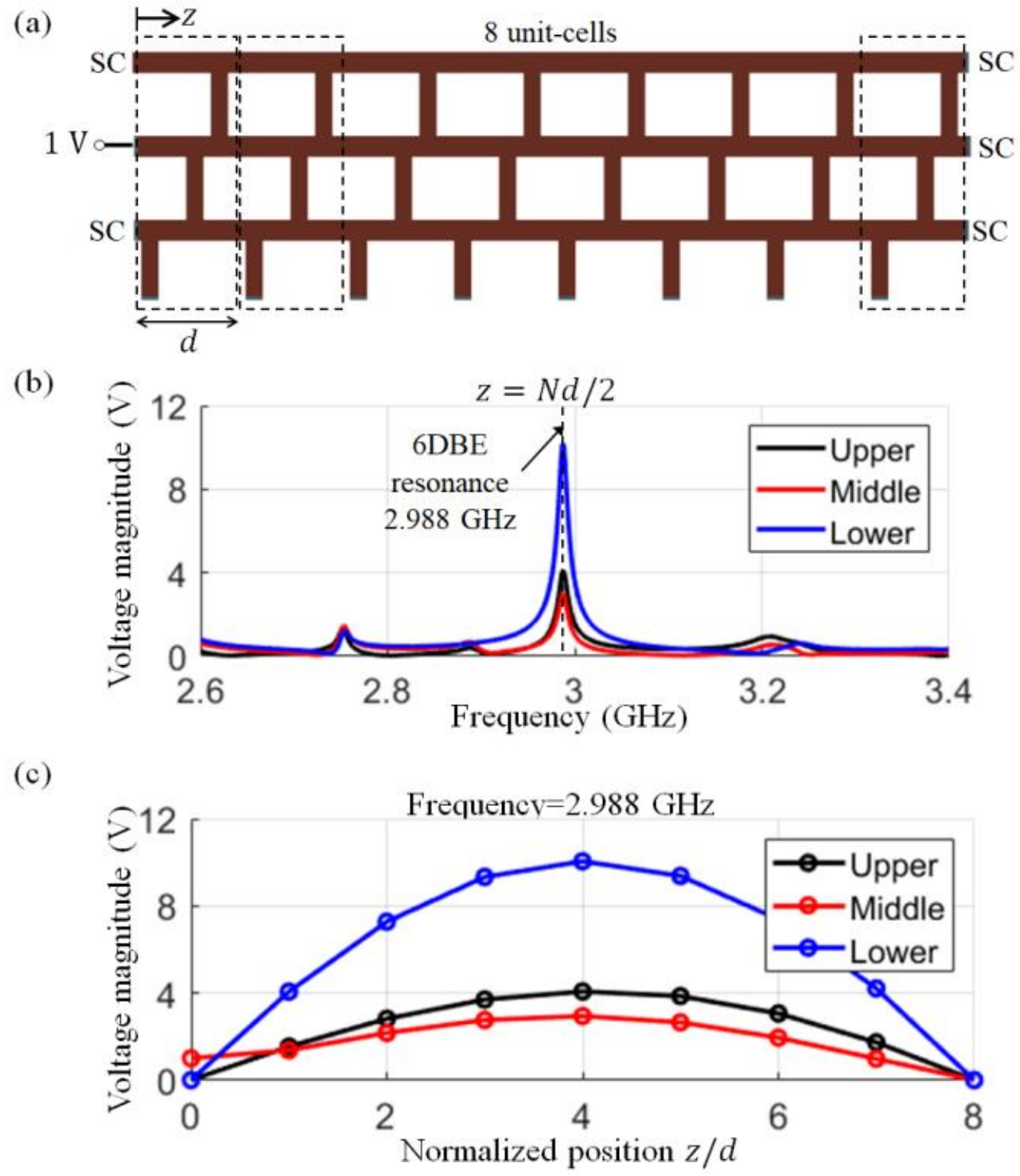

Fig. 2 (a) A resonance cavity formed by cascading 8 unit cells (dielectric and underneath ground plane not shown). The left and right terminations are shorted to the ground except for the middle line at the left end of the structure. We have applied a phasor voltage of 1 V as input to the resonator. (b) Frequency response of the voltage at the middle of the cavity where a resonance peak is observed at a frequency close to the 6DBE frequency. (c) Voltage profile across the structure at the 6DBE resonance frequency showing the voltage distribution at each period of the three TLs .

The 6 × 6 *S*-parameters matrix was acquired by only using 2 ports of the VNA at a time as illustrated in the following way. Port 1 of the VNA is connected to port *m* of the unit cell while, port 2 of the VNA is connected to port *n* of the unit cell while the four other ports are terminated by a 50 Ω load, to measure *S*(*m,n*) (as well as *S(n,m)*). The obtained S-parameters are smoothed out over frequency using Matlab to eliminate the effect of random noise. Once the 6 × 6 *S*-parameters matrix is constructed, we obtained the 6 × 6 transfer matrix $\underline{\mathbf{T}}$ from which the Bloch wavenumbers are solved for (as described previously). The wavenumbers for both the measured and simulated results are plotted versus frequency to obtain the dispersion diagram shown in Fig. 3(b). These results show some perturbation and deviation from the ideal degeneracy condition (flat dispersion diagram at 6DBE frequency) in Fig. 1(b). The perturbation is due to the ohmic, dielectric and radiation losses of the waveguide, and to tolerances in the fabrication..

Since an EPD is very sensitive to perturbations, we use a figure of merit to describe "how far" we are from an ideal EPD, based on the concept of *coalescence parameter* or *hyperdistance*, $C(\omega)$ , *between the six eigenvectors* of the transfer matrix of one-unit cell $\underline{\mathbf{T}}$ defined in [5]. The matrix is here obtained via measurements and full-wave simulations. The coalescence parameter that defines the closeness to the 6DBE is defined as

$$C = \frac{1}{15} \sum_{\substack{m=1,n=1 \\ m \neq n}}^{6} \left| \sin\left(\theta_{mn}\right) \right|, \quad \cos\left(\theta_{mn}\right) = \frac{\left| \left( \mathbf{\Psi}_m, \mathbf{\Psi}_n \right) \right|}{\left\| \mathbf{\Psi}_m \right\| \left\| \mathbf{\Psi}_n \right\|} \tag{2}$$

where $\theta_{mn}$ represents the angle between two eigenvectors $\mathbf{\Psi}_m$ and $\mathbf{\Psi}_n$ in a six-dimensional complex vector space, with norms $\left\| \mathbf{\Psi}_m \right\|$ and $\left\| \mathbf{\Psi}_n \right\|$, and $\left( \mathbf{\Psi}_m, \mathbf{\Psi}_n \right)$ as their inner product (involving complex comjugation). This coalescence parameter helps determine the closeness of a system to having a 6DBE and distinguishing this 6DBE from other EPDs with different orders (like the regular band edge or the DBE). The coalescence parameter presented here is a very convenient figure to assess if an EPD occurs in reality, when losses and other perturbations are present. The perfect 6DBE that exists only mathematically would provide a $C = 0$. However a system can still preserve the full degeneracy properties when the eigenvectors are very close to each other, i.e., in a neighborhood of the 6DBE. The coalescence factor, *C*, is plotted in Fig 3(c) as a function of frequency, accounting for all dissipative and radiative losses, showing a good agreement between the two results based on measurements and full-wave simulations, besides a frequency shift. It can be seen that $C$ has the minimum value in the vicinity of the 6DBE frequency ( $f = f_d$ ) as expected. In an ideal lossless case it is expected that $C \to 0$ at the exact EPD.

The occurrence of 6DBE implies that the system matrix is non-diagonalizable because the system has 6 repeated eigenvalues and one eigenvector. In other words, at the 6DBE, only one polarizations state $\mathbf{\Psi}_e$ is the eigenvector of the system, the other polarizations states can be described by generalized eigenvectors. This implies that the geometric multiplicity of the degenerate eigenvalue is equal to 1 while its algebraic multiplicity is equal to 6, hence the transfer matrix $\underline{\mathbf{T}}$ is not diagonalizable and it is similar to a Jordan block matrix of 6 × 6 dimension. Therefore, at the 6DBE, the transfer matrix $\underline{\mathbf{T}}$ is represented as

$$\underline{\mathbf{T}} = \underline{\mathbf{V}} \underline{\mathbf{\Lambda}}_{\mathrm{J}} \underline{\mathbf{V}}^{-1} , \tag{3}$$

where $\underline{\mathbf{\Lambda}}_{\mathrm{J}}$ is the Jordan matrix

$$\underline{\mathbf{\Lambda}}_{\mathrm{J}} = \begin{bmatrix} \zeta_e & 1 & 0 & 0 & 0 & 0 \\ 0 & \zeta_e & 1 & 0 & 0 & 0 \\ 0 & 0 & \zeta_e & 1 & 0 & 0 \\ 0 & 0 & 0 & \zeta_e & 1 & 0 \\ 0 & 0 & 0 & 0 & \zeta_e & 1 \\ 0 & 0 & 0 & 0 & 0 & \zeta_e \end{bmatrix}, \tag{4}$$

and $\zeta_e = e^{-jk_e d}$ . The similarity transformation matrix $\underline{\mathbf{V}}$ is composed of one degenerate eigenvector and five generalized eigenvectors, associated with the eigenvalue $\zeta_e$ .

In practice, due to the fabrication tolerances and losses of the structure, the system can not operate exactly at the 6DBE but rather very close to it. When the system is close to the EPD, the system matrix is very close to be in a Jordan block form. At this case one can measure how close the system matrix is to Jordan block through measuring how the eigenvector of the system are close to each other using the prementioned concept of the coalasance parameter as done in Fig. 3.It is important to point out that the shift in the 6DBE frequency in the dispersion diagrams obtained from the measurement and full-wave simulation is mainly due to the extra lengths accounted in the SMA connectors. Deemebeding techniques, similar to the ones in [15], [27], can be applied to remove the effect of the SMA connectors on the measurement. Note that the dispersion obtained from measurement in Fig. 3(b) preserves the shape of 6DBE although deembeding was not performed, and this shows that some tolerances can be fully accepted in the design. We confirmed that the frequency shift is is mainly due to the described effect by performing another full-wave simulation for unit cell in Fig. 1(a), for a similar unit cell but after adding 1mm extension of TL at each port (hence the unit-cell period is extended by 2mm). We show in Fig. 3(d) a comparison between the two dispersion diagrams that shows that the length extension does not impact the occurrence of 6DBE but rather shifts its frequency. Therefore, we concluded that the SMA connectors extra lengths might lower the frequency at which the 6DBE occurs as observed in Fig. 3(b). This result also shows a possible way to tune the 6DBE frequency.

## V. Conclusion

A physical realization of a three-way periodic waveguide exhibiting a sixth-order degeneracy (6DBE) is demonstrated using microstrip technology. The 6DBE is an exceptional point of order six where six eigenstates coalesce onto a single degenerate one. Both theoretical and experimental verifications are provided. Remarkable physical properties may arise due to this strong sixth-order exceptional point of degeneracy including increased quality factor, high density of states, and sensitivity which can lead to novel designs for microwave and optical pulse generation [19], microwave and millimeter wave oscillators [23], [24], [26], [28], low threshold lasers [17], [29], short delay lines with significant group delay, filters, and ultra-sensitive sensors. The first fabrication and experimental demonstration of a DBE oscillator in microstrip technology was shown in [30], exhibiting very robust oscillation behavior. The 6DBE may lead to analogous or even superior performance and it should be further investigated.

## Acknowledgment

This material is based upon work supported by the National Science Foundation under award NSF ECCS-1711975 and by the Air Force Office of Scientific Research under award number FA9550-15-1-0280.

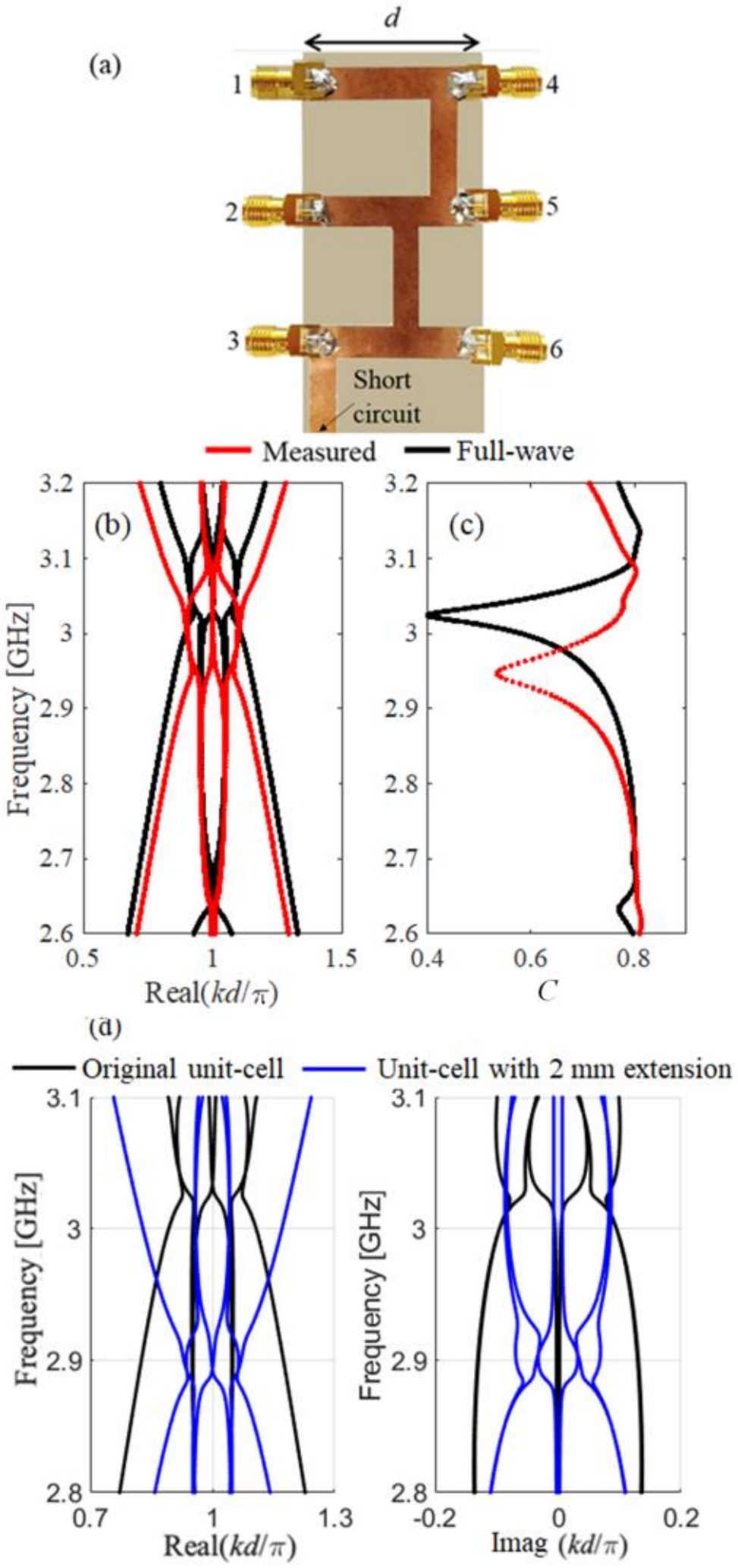

Fig. 3 (a) Fabricated unit cell of the proposed periodic three-way waveguide implemented in microstrip technology (the ground plane is below the dielectr substrate (light brown). (b) Bloch wavenumbers dispersion based on measurement (red dots) versus the one obtained via Method of Moments full-wave electromagnetic simulation (solid black). Both measurement and simulation are based on a six-port evaluation of the scattering parameters of the unit cell. Six modes coalescing are clearly visible, though the coalescence is not perfect because of losses and manufacturing imperfections. (c) Coalescence factor, $C$ in linear scale versus frequency around the 6DBE, as a measure of closeness to the 6DBE, for the full-wave simulation results, in comparison to the measured results. (d) Tunability of 6DBE showing dispersion diagrams obtained from full-wave simulation of unit cell in Fig. 1(a) (black) and same unit cell after adding 1mm extension of transmission line at each port (period is extended by 2mm). The comparison shows that 6DBE frequency is very sensitive to any